\documentclass[11pt,american]{article}
\usepackage[T1]{fontenc}
\usepackage[latin9]{inputenc}
\usepackage[letterpaper]{geometry}
\usepackage{graphicx}
\usepackage{setspace}
\usepackage{subscript}

\makeatletter
\newcommand{\lyxaddress}[1]{
\par {\raggedright #1
\vspace{1.4em}
\noindent\par}
}
\newenvironment{lyxlist}[1]
{\begin{list}{}
{\settowidth{\labelwidth}{#1}
 \setlength{\leftmargin}{\labelwidth}
 \addtolength{\leftmargin}{\labelsep}
 }}
{\end{list}}

\makeatother

\usepackage{babel}
\begin{document}

\title{\textbf{Functional Magnetic Resonance Imaging: a study of malnourished
rats }}

\author{R. Martin%
\thanks{{*}Corresponding author: M. Sc. Rodrigo A. Martin, email: mriuami@gmail.com%
}, R. Godinez, A. O. Rodriguez}

\maketitle

\lyxaddress{\begin{center}
Department of Electrical Engineering, Universidad Autonoma Metropolitana
Iztapalapa, Av. San Rafael Atlixco 186, Mexico, DF, 09340, Mexico
\par\end{center}}
\begin{abstract}
Malnutrition is a main public health problem in developing countries.
Incidence is increasing and the mortality rate is still high. Malnutrition
can leads mayor problems that can be irreversible if it is present
before brain development is completed. We used BOLD Functional Magnetic
Resonance Imaging to investigate the regions of brain activity in
malnourished rats. The food competition method was applied to a rat
model to provoke malnutrition during lactation. The weight increase
is delayed even if there is plenty milk available. To localize those
regions of activity resulting from the trigeminal nerve stimulation,
the vibrissae-barrel axis was employed due to the functional and morphological
correlation between the vibrissae and the barrels. BOLD response changes
caused by the trigeminal nerve stimulation on brain activity of malnourished
and control rats were obtained at 7T. Results showed a major neuronal
activity in malnourished rats on regions like cerebellum, somatosensorial
cortex, hippocampus, and hypothalamus. This is the first study in
malnourished rats and illustrates BOLD activation in various brain
structures.
\end{abstract}

\section{Introduction}

The discovery of the cerebral processes of cerebral maturation in
mammals, have held the opportunity to investigate that exists a certain
vulnerability in the cerebral development when there is a malnourished
problem, causing cerebral damage {[}1{]}. Among the important is that
we can see damage at both morphological and neurotransmitter levels
{[}1{]}. 

Different studies in malnourished animal models, such as post-mortem,
histological and imaging have helped to discover some morphological
changes including a decrease in the prenatal brain size {[}2{]}, the
postnatal brain size {[}3{]}, the hippocampal neural cells {[}4{]},
the granular volume at the cerebellar cortex {[}5{]}, abnormalities
in the dendritic spines at the cerebellar cortex {[}6{]}, reduction
in the myelin content {[}7{]}, and the protein synthesis of myelin
{[}8{]}, among others.

Other specific consequences in brain function are principally alterations
in neurotransmitters. Reduction of serotonin levels {[}9{]} {[}10{]},
increase of GABA concentrations at the hippocampus {[}11{]}, reduction
in the cholinergic cells {[}12{]}, increase of the AChE at the hippocampus
{[}13{]}, decrease in the segregation of dopamine {[}14{]} {[}15{]},
have been found such as hypothalamus, hippocampus, and cerebral cortex,
areas related to memory and learning. 

Almost all of the functional in vivo records have been made with electroencephalography
methods, using both, conventional and invasive {[}1{]}. Electroencephalographic
methods can give us many noisy signals that can lead into a malinterpretation
of results if there is not an expertise. The quality of the electroencephalographic
signals depends on the type and number of electrodes employed, that
can carry out problems at the moment of placing them. We have to consider
that at the moment of inserting the electrodes we can damage some
brain tissues. Magnetic Resonance Imaging (MRI) is a powerful imaging
tool which has many advantages in comparison to other imaging techniques.
MRI is a non-ionizing technique, able to produce anatomical and functional
information. Image contrast depends on the tissue intrinsic characteristics
making it ideal to study brain activity. In this work, we used functional
Magnetic Resonance Imaging (fMRI), which provides a method for mapping
brain functional activity based on the blood oxygen level-dependent
effect (BOLD). The BOLD MRI technique, allows indirectly to study
the whole brain activation, in an animal model under a malnourished
program.

As far as we know, this is the first fMRI study on malnutrition. To
provoke malnutrition, the food competition method was applied {[}16{]}.
This method consists in inducing malnutrition during lactation through
food competition. A large number of pups cannot be sufficiently fed
by one nursing mother. Then, a delay in the weight increase is observed
even if there is plenty milk available. We chose the stimulation of
the vibrissae-barrel axis because it is suitable for studying structure,
function, development and plasticity within the somatosensory cortex,
due to the functional and morphological correlation between the vibrissae
and the barrels {[}17{]}.

\section{Materials and methods}

\subsection{Animal preparation}

All animal procedures were performed according to the federal guidelines
of the Animal Care and approved by the local authority. Twelve male
Wistar rats (provided by the closed colony of breeding of the Division
of Biological and Health Sciences of the Autonomous Metropolitan University,
Mexico) were used. The animals were kept in an environment at a temperature
of 22 - 25 \textsuperscript{0}C, with 45\% relative humidity. Animals
were kept under 12 hours controlled light-darkness cycles. Animals
were divided into two groups according to nutritional manipulation:
control and experimental groups. Both groups consisted of 6 rats each
(aged 18 to 21 days). To provoke malnutrition, the food competition
method was applied {[}17{]}. Control rats weighted 41.94$\pm$4.9g
and malnourished rats weighted 29.09$\pm$3.3g. The weights of the
experimental group correspond to a second degree malnourished level,
according to {[}16{]}. This setup can be seen in Figure 1.\medskip{}
\begin{minipage}[t]{1\columnwidth}%
\begin{lyxlist}{00.00.0000}
\begin{singlespace}
\item [{\includegraphics[scale=0.6]{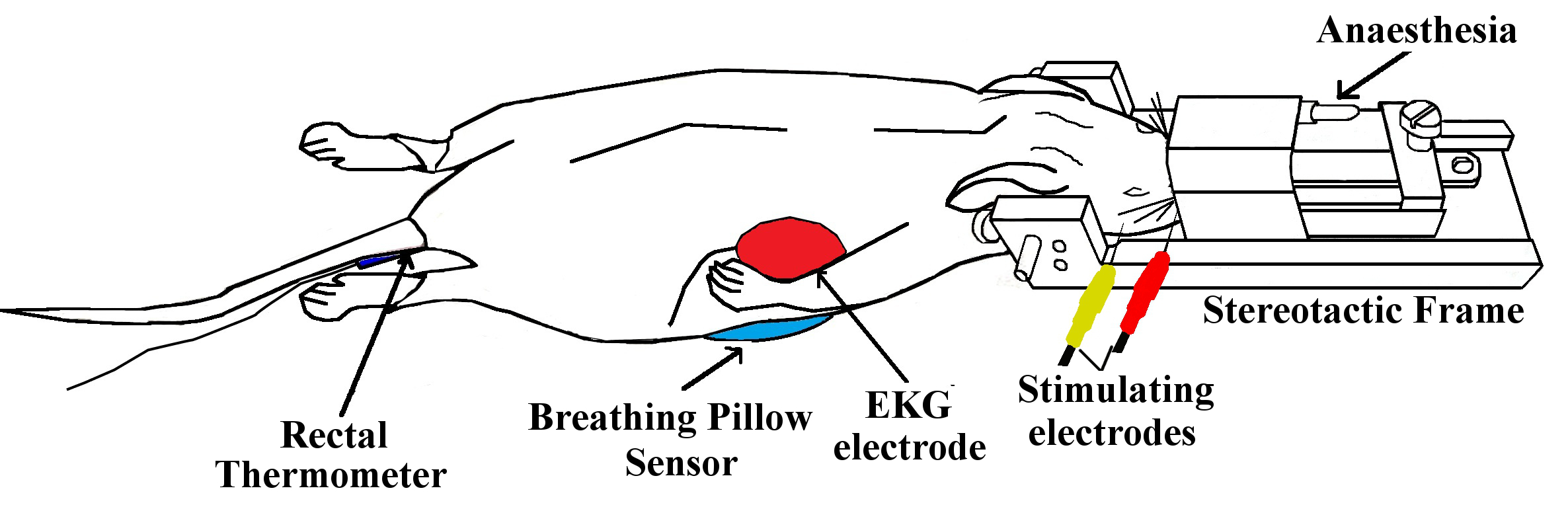}}]~\end{singlespace}

\end{lyxlist}
\begin{singlespace}
\begin{center}
{\scriptsize Figure 1. Animal Setup employed for all BOLD experiments.
Stimulating electrodes were placed on the whiskers (anode) and in
the masticatory muscles (cathode). The anesthesia was delivered through
the tube attached to the stereotactic frame-like. To minimise motion
artefacts, the rat's head and ears were fixed to an Agilent stereotactic
frame. This setup was also used to administer anesthesia at 1\% of
isoflurane with 2 l/min of O2 during all the fMRI experiment. EKG,
breathing monitoring, and rat's temperature control (37\textsuperscript{0}C
$\pm$ 1\textsuperscript{0}C) were done with a small animal monitoring
and gating system (Model 1025, SA Instruments, NY, USA).}
\par\end{center}\end{singlespace}
\end{minipage}

\subsection{Trigeminal nerve stimulation}

The trigeminal nerve (the fifth cranial nerve, also called the fifth
nerve, or simply CNV or CN5) contains both sensory and motor fibers.
It is responsible for sensation in the face and certain motor functions
such as biting, chewing, and swallowing. Sensory information from
the face and body is processed by parallel pathways in the central
nervous system. In animal models, more specific in rodents, it has
been used as an attractive model the whisker sensory system for studying
this nerve. This structure helps us to study and understand structure,
function, development, and plasticity within the somatosensory cortex
{[}17{]}.

\subsection{fMRI Acquisition and data processing}

All experiments were performed on a 7T/21cm Agilent system (Agilent
Technologies, Inc, Palo Alto, CA) equipped with DirectDrive technology
and a transceiver 16-rung birdcage coil (12 cm long and a 6 cm diameter).
Rat's brain images were acquired using a standard gradient echo sequence
and the following parameters: TR/TE=107.82/3.8 ms, Flip angle=200,
FOV=30x30mm, matrix size= 128x128, thickness=0.3mm, and NEX=1, thus
a spatial resolution of 0.23 x 0.23 x 0.3 mm3 was achieved. Gradient-echo
BOLD fMRI has reasonably high spatiotemporal resolution can be routinely
performed on this type of animal models at high field.

Images for BOLD fMRI were taken for both control and experimental
groups. The left trigeminal nerve was stimulated using percutaneously
inserted stainless steel electrodes, whose cathode was positioned
in the whiskers and the anode was inserted in the masticatory muscles.
The trigeminal nerve was stimulated using a stimulator with 10 ms
constant current pulses (500 mV and 2 mA) applied every second (Grass
S-48 Stimulator, Grass Technologies, RI, USA) and at 1Hz. 60s OFF
was alternated with 60s ON periods {[}17{]}.

To determine the regions of interest in the somatosensory cortex where
neuronal activity is expected to happen, all brain images were digitally
processed using the toolbox SPMMouse {[}18{]}. Data were overlaid
onto anatomical images acquired in the same image plane. For comparison
between control and experimental somatosensory cortex we used a p
< 0.005, and for the analysis in the experimental group a p < 0.05
was used, for estimating the SPM results.

\section{Results }

\subsection{fMRI maps}

To study the possible areas of activation, representative activation
maps of stimulus-correlated intensity changes in the rat brain of
both groups were obtained. Fig. 2 shows five different brain regions
from the olfactory bulb to the cerebellum on both groups, so we can
appreciate different activation between groups. Schematics diagrams
taken from Paxino's rat brain atlas were added to facilitate localization
of the coronal cuts in the rat's brain {[}19{]}. Under normal conditions,
the expected responses are similar to the response of the control
group, as shown in Figure 2. 

\medskip{}

\begin{minipage}[t]{1\columnwidth}%
\begin{center}
\includegraphics[scale=0.5]{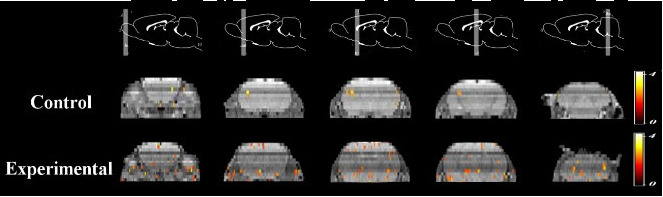}
\par\end{center}

\begin{singlespace}
{\scriptsize Figure 2. Coronal brain maps of BOLD intensity of one
rat of each group showing activation from the olfactory bulb to cerebellum.
From left to right, we see areas corresponding from olfactory bulb
to the cerebellum. Notice that in the experimental group exist random
pattern activation in comparison with the control group (expected
response).}\end{singlespace}
\end{minipage}

This is an important activation only in somatosensory cortex per se.
But in the case of a malnourished brain, it can be appreciated responses
from other brain structures, such as hippocampus, hypothalamus, and
cerebellum besides somatosensoy cortex. In general, all BOLD fMRI
maps of malnourished rats show more activation than those in the control
group in the entire brain. However, the brain may be regulated by
a fine mechanism of excitatory and inhibitory entries at the interneuron
level {[}20{]}. Our results indicate an abnormal increment in the
electrical activity among different regions of the cerebral tissue.
This is probably the result of a decrease on the neural inhibitory
discharges of the inhibitory postsynaptic potentials (IPSP) and/or
an increase in the excitatory postsynaptic potentials (EPSP) {[}20{]}.
Results reported by Segura and collaborators {[}21{]} demonstrated
the existence of a demyelinating process in the peripheral nervous
system in malnourished rats and, that a similar effect is happening
at the encephalic level. Malnutrition probably is affecting the inhibitory
interneurons, so that their demyelination processes cause a decrease
in the IPSP, and then increasing the neuronal activity in a random
way as depicted in Figure 2. From our results, it is not yet possible
to determine if abnormal activation from different brain regions is
a consequence of brain's tissue damage or these activations are the
result from adaptive process of neural plasticity. The changes are
so dramatic that suggest the existence of preexistent synaptic pathways
that are not normally expressed, while in a brain of a malnourished
subject these pathways are unmasked.

\subsection{BOLD response}

Control and experimental data at the somatosensory cortex were then
fitted to a non-linear regression over time and shown in Figure 3.
Nonlinearity of the BOLD response is observed for both groups. There
is great similarity in the pattern of BOLD response for both groups.
These results showed a great concordance with results already reported
by Silva and Korestky {[}22{]}. The BOLD responses experiment a similar
time delay for both groups and reach their maximum value at 5 seconds
during the stimulation. Experimental group results suggest larger
oxygen consumption when compared with the control group. Fig. 3c)
shows the averaged BOLD signal intensity time course for control and
experimental groups were computed with 6 rats: 30 seconds per rat.
A sign change can be appreciated between the regions I and II of Fig.
3.c), roughly after 12 s for both cases. 

\medskip{}

\begin{minipage}[t]{1\columnwidth}%
\begin{center}
\includegraphics[scale=0.8]{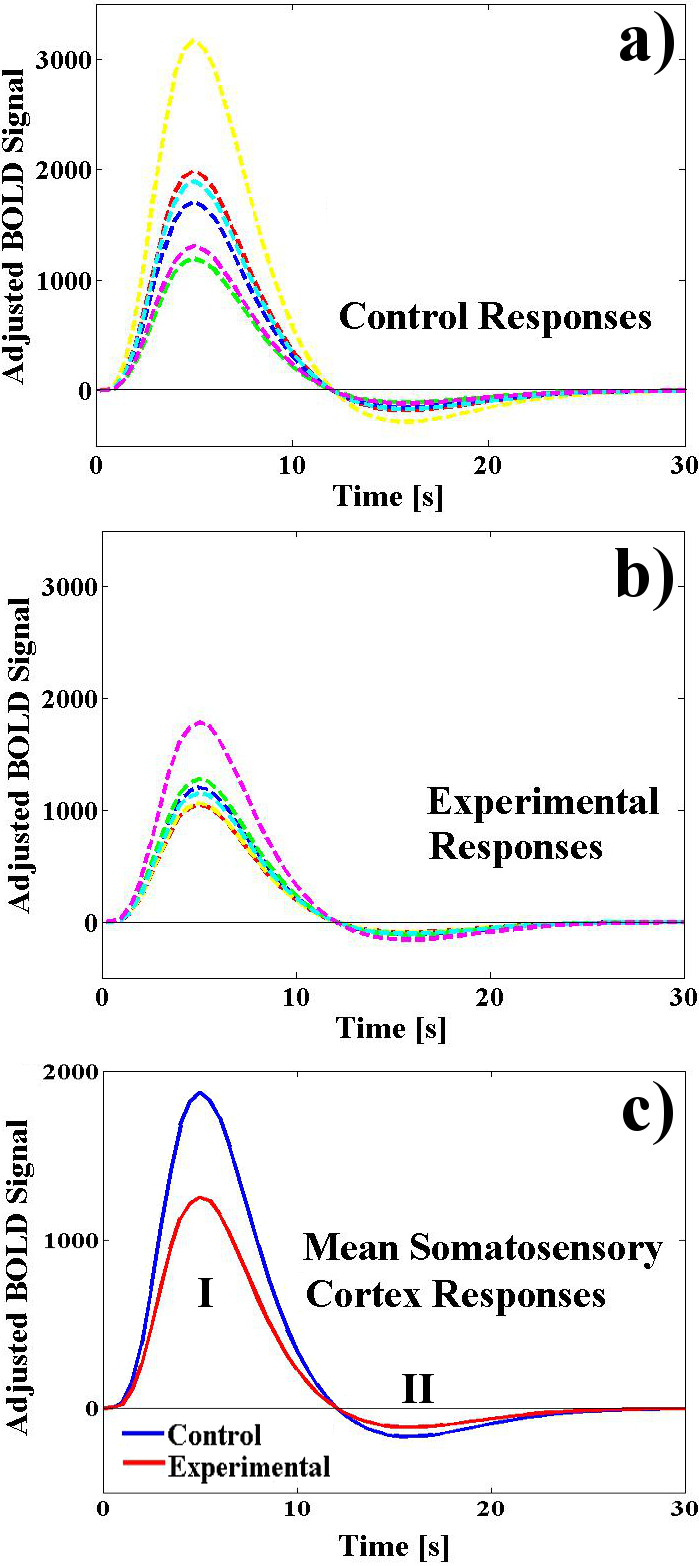}
\par\end{center}

\begin{singlespace}
\begin{flushleft}
{\scriptsize Figure 3. Adjusted BOLD signal from the somatosensory
cortex obtained from a) control group, b) experimental group, and
c) the mean response from each group of 6 rats each.} 
\par\end{flushleft}\end{singlespace}
\end{minipage}

\medskip{}

This has been already reported as a negative BOLD response and it
is caused by inhibition mechanisms {[}23{]}. A t student test was
run to investigate statistical independence of the measurements of
the response amplitude. At the 0.05 level, the difference of the population
means (mean\textsubscript{control}=1879.39 \& mean\textsubscript{exp}=1254.34)
is statistically different. In order to quantify the BOLD response
of the malnourished rats, the method reported in {[}24{]} was used.
Then, the response amplitudes were calculated from the peak of BOLD
response intensity. Additionally, the response integral was defined
as the product of the amplitude and the full width at half maximum
(FWHM) for both groups of rats the response integral values are 5638.18
and 3763.01 for the control and malnourished rats, respectively. Also
a simple trapezoid-integration under the positive curve was made;
the results were 20460.15 and 13708.85 for control and malnourished
rats, respectively. This parameter indicates a clear increment in
the neuronal activity for the malnourished rats. 

We also analyzed responses in the experimental group, in Figure 4,
we can see BOLD responses for different structures that were activated,
and we can see responses from cerebellum, hypothalamus, hippocampus,
and somatosensory cortex. We made the same measurements of area under
the positive curve of the mean responses and the results obtained
by simple trapezoid-rule integration were: cerebellum$\approx$$ $21625.49,
hypothalamus$\approx$$ $15606.45, hippocampus$\approx$$ $15272.86,
and somatosensory cortex$\approx$$ $13708.85. 

\medskip{}

\begin{minipage}[t]{1\columnwidth}%
\begin{center}
\includegraphics[scale=1.1]{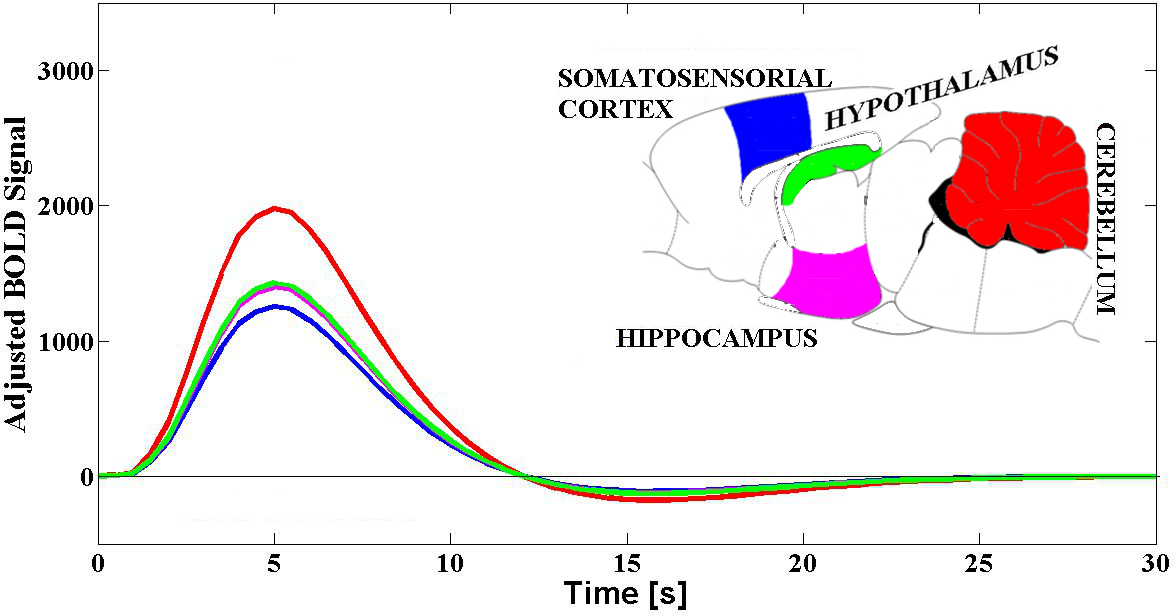}
\par\end{center}

\begin{singlespace}
{\scriptsize Figure 4. Adjusted BOLD signals from different brain
areas from experimental (malnourished) group. Also there is represented
the mean BOLD signal from each brain areas. All figures show same
maximum intensity scale so we can notice the differences between structures.
The lower figure show areas from which this signals were taken.}\end{singlespace}
\end{minipage}

\medskip{}
A similar behavior was obtained with the FWHM method, described for
the comparison between control and experimental groups; in the malnourished
analysis the results were cerebellum$\approx$$ $5936.09, hypothalamus$\approx$$ $4283.9,
hippocampus$\approx$$ $4192.33, and somatosensory cortex$\approx$$ $3763.01.
A greater area at the cerebellum region can be appreciated, hypothalamus
and hippocampus have a very similar behavior, and somatosensory cortex
area has a lower area in comparison to the cerebellum one, and larger
area means a greater activity in the region. Our results agree with
the results obtained with invasive techniques and demonstrated that
cerebellum is one of the structures more affected by malnutrition
as demonstrated by Hillman {[}5{]}. Other structures related to learning
and memory processes are also affected because of the malnutrition
process, such as hippocampus and hypothalamus {[}2-14{]}. Finally
a three-dimensional rendering process was computed to observe anatomical
differences between both groups. In Fig. 5 we can clearly observe
significant differences in brain structure, mainly at the cerebellum
development. Also we can notice a big difference at the development
of both brain hemispheres.

\medskip{}

\begin{minipage}[t]{1\columnwidth}%
\begin{singlespace}
\begin{center}
\includegraphics[scale=0.3]{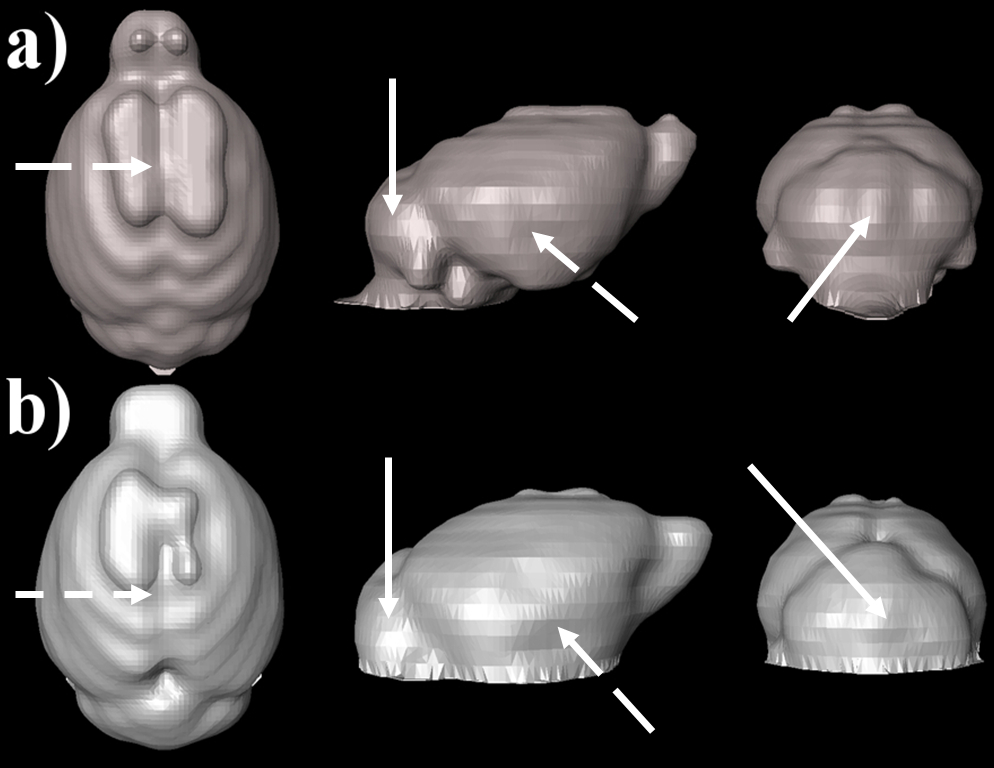}
\par\end{center}
\end{singlespace}

\begin{singlespace}
{\scriptsize Figure 5. 3D reconstruction from the anatomical images
obtained from the a) top row, control and b) bottom row, experimental
group. We can see an important difference between both brains, at
the cerebellum (solid line arrow), in which a notorious low development
is present. Also an irregular development from both cerebral hemispheres
(dashed line arrow) can be noticed in the experimental brain.}\end{singlespace}
\end{minipage}

\section{Discussion}

This fMRI study is the first performed in rats and illustrates a similar
pattern of BOLD responses between malnourished and control rats. Negative
BOLD responses can be also appreciated for the two groups of rats
caused by inhibition mechanisms. Vascular nonlinearities are the major
contribution to the observed nonlinearities and they are neuronal
in origin. Our results make a pathway to continue study this health
problem and its consequences in brain function, because we cannot
still say if the resulting changes in the brain correspond totally
to a brain damage or if there is a process of adaptive plasticity.

Results obtained were not the expected ones, although it has been
described brain damages by malnutrition in structures like cerebellum,
hypothalamus, and hippocampus; according to the experimental protocol
used. We only expected limited alterations at the somatosensoy cortex
at first, and no significant changes in the function of other structures.
The brain changes observed with this experiment may imply that the
mechanisms of neurotransmitters like GABA (inhibitory) or Glutamate
(excitatory) are affected, but others neurotransmitters like ACh,
Dopamine, or Noradrenaline may be not working as they should.

Also, our results may imply that exist problems with the hypothalamic
pathway with other structures. These pathways are shown in the brain
scheme of Fig. 5. These problems may be involved with the neurotransmitters
cycle.

\section{Conclusion}

We used BOLD fMRI and trigeminal nerve stimulation to study the BOLD
changes during sensory stimulation in malnourished rats. This study
is the first performed in malnourished rats and illustrates BOLD activation
in various brain areas, specially cerebellum is the structure with
more activation. Further investigation should be carried out to determine
whether plasticity or demyelination, or even combinations of both
are responsible for the activation of the brain areas reported in
this work. These results may pave the way into the treatment of this
health problem and other possible rehabilitation procedure.

\section*{Acknowledgments }

We would like to acknowledge financial funding from CONACYT-Mexico
under grant no. 166404 and a Ph. D. scholarship.

\section*{References}
\begin{enumerate}
\item \begin{flushleft}
M. Medina, et. al. Consequences of Malnutrition. World Federation
of Neurology. Seminars in Clinical Neurology. Vol 6. 2007. pp 1-36.
\par\end{flushleft}
\item \begin{flushleft}
Zeman FJ, et. al. Teratog Carcinog Mutagen, 6: 339-347. 1986.
\par\end{flushleft}
\item \begin{flushleft}
Fukuda MT, et. al. Behav Brain Res, 133: 271-277. 2002.
\par\end{flushleft}
\item \begin{flushleft}
Lister JP, et. al. Hippocampus, 15: 393-403. 2005.
\par\end{flushleft}
\item \begin{flushleft}
Hillman DE, Chen S. Neuroscience, 6: 1249-1262. 1981
\par\end{flushleft}
\item \begin{flushleft}
Benitez-Bribiesca L, et. al. Pediatrics, 104: e21. 1999
\par\end{flushleft}
\item \begin{flushleft}
Reddy PV, et. al. Brain Res,161: 227-235 . 1979.
\par\end{flushleft}
\item \begin{flushleft}
Montanha-Rojas EA, et. al. Nutr Neurosci, 8: 49-56. 2005.
\par\end{flushleft}
\item \begin{flushleft}
Mazer C, et. al. Brain Res, 760: 68-73. 1997.
\par\end{flushleft}
\item \begin{flushleft}
Chen JC, et. al. Int J Dev Neurosci, 15: 257-263. 1997.
\par\end{flushleft}
\item \begin{flushleft}
Chang YM, et. al. Nutr Neurosci, 6: 263-267. 2003.
\par\end{flushleft}
\item \begin{flushleft}
Nakagawasai O. Yakugaku Zasshi, 125: 549-554. 2005.
\par\end{flushleft}
\item \begin{flushleft}
Cermak JM, et. al. Dev Neurosci, 21: 94-104. 1999.
\par\end{flushleft}
\item \begin{flushleft}
Zimmer L, et. al. Neurosci Lett, 284: 25-28. 2000.
\par\end{flushleft}
\item \begin{flushleft}
Chalon S, et. al. Lipids, 36: 937-944. 2001.
\par\end{flushleft}
\item \begin{flushleft}
Ortiz R. et al. Med. Sci. Res. 24:843. 1996
\par\end{flushleft}
\item \begin{flushleft}
Just N. et al. Mag. Res. Imaging. 28:1143-1151. 2010.
\par\end{flushleft}
\item \begin{flushleft}
Sawiak SJ. et. al. 17th ISMRM. 2009:1086
\par\end{flushleft}
\item \begin{flushleft}
Paxinos G, Watson Ch. The Rat Brain in Stereotaxic Coordinates, 4th
Ed. Academic Press, 1998
\par\end{flushleft}
\item \begin{flushleft}
Kandel, ER. Principles of Neural Science, 4th Ed. McGraw-Hill, New
York, USA. 2000.
\par\end{flushleft}
\item \begin{flushleft}
Segura B. et. al. Neurosc. Lett. 2004;354:181
\par\end{flushleft}
\item \begin{flushleft}
Silva AC, Koretsky AP. PNAS. 2002;99:15182
\par\end{flushleft}
\item \begin{flushleft}
Vandervliet E. et al. ESMRMB. 2006:624.
\par\end{flushleft}
\item \begin{flushleft}
Yesilyrut B. et al. MRI. 2008;26:85.
\par\end{flushleft}
\end{enumerate}
\begin{singlespace}
\end{singlespace}

\end{document}